\documentclass{PoS}
\usepackage{cite}

\def\samurai{{{\sc Samurai}}}

\def\C++{{{\sc c++}}}

\newcommand{\beq}{\begin{equation}}
\newcommand{\eeq}{\end{equation}}
\newcommand{\bqa}{\begin{eqnarray}}
\newcommand{\eqa}{\end{eqnarray}}
\newcommand{\bite}{\begin{itemize}}
\newcommand{\eite}{\end{itemize}}

\def\db#1{ D_{#1}}

\title{The Integrand Reduction of One- and Two-Loop Scattering Amplitudes}

\ShortTitle{Integrand Reduction}

\author{\speaker{Pierpaolo Mastrolia} \\
Max-Planck Insitut f\"ur Physik, M\"unchen, Germany\\
Dipartimento di Fisica e Astronomia, Universit\`a di Padova, and INFN                                    
Sezione di Padova, Italy\\
 E-mail: \email{ppaolo@mpp.mpg.de}}

\author{Edoardo Mirabella \\
Max-Planck Insitut f\"ur Physik, M\"unchen, Germany\\ 
 E-mail: \email{mirabell@mppmu.mpg.de}}

\author{Giovanni Ossola\\
 New York City College of Technology, City University of New York, USA\\
 Graduate School and University Center, City University of New York, USA \\ 
 E-mail: \email{gossola@citytech.cuny.edu}}
 
\author{Tiziano Peraro \\
Max-Planck Insitut f\"ur Physik, M\"unchen, Germany\\ 
 E-mail: \email{peraro@mppmu.mpg.de}}

\author{ Hans van Deurzen\\
Max-Planck Insitut f\"ur Physik, M\"unchen, Germany\\ 
 E-mail: \email{hdeurzen@mppmu.mpg.de}}

\abstract{The integrand-level methods for the reduction of scattering amplitudes are well-established techniques, which have already proven their effectiveness in several applications at one-loop. In addition to the automation and refinement of tools for one-loop calculations, during the past year we observed very interesting progress in developing new techniques for amplitudes at two- and higher-loops, based on similar principles. In this presentation, we review the main features of integrand-level approaches with a particular focus on algebraic techniques, such as Laurent series expansion which we used to improve the one-loop reduction, and multivariate polynomial division which unveils the structure of multi-loop amplitudes. }

\FullConference{Loops and Legs in Quantum Field Theory - 11th DESY Workshop on Elementary Particle Physics,\\
		April 15-20, 2012\\
		Wernigerode, Germany}

\begin{document}

\section{Overview}

Scattering amplitudes are analytic functions of the momenta of the particles involved and they can be studied by exploring their singularity structure~\cite{Bern:1994zx,Cachazo:2004kj}.
The investigation of the residues at the poles, which correspond to particles going on their mass-shell,  led to the
discovery of new important relations. The BCFW  recurrence relation \cite{Britto:2004ap},
its link to the leading singularity of one-loop amplitudes \cite{Britto:2004nc},
and the OPP integrand-decomposition formula for  one-loop
integrals \cite{Ossola:2006us} have shown the underlying simplicity beneath 
the rich mathematical structure of quantum field theory.
Moreover they provided the theoretical framework to develop efficient computational techniques at the next-to-leading order in perturbation theory \cite{Berger:2008sj,Giele:2008bc,Bevilacqua:2011xh, Hirschi:2011pa, Cullen:2011ac,Agrawal:2011tm,Cascioli:2011va,Badger:2012pg}.

The integrand reduction methods, originally developed for one-loop diagrams~\cite{Ossola:2006us}, 
use the singularity structure of the integrands to decompose the (integrated) amplitudes in terms of Master Integrals (MI's). 
The multi-particle pole expansion of the integrand 
is equivalent to the decomposition of the numerator in terms of 
products of denominators, multiplied by polynomials which
correspond to the residues at the multiple-cuts.

The parametric form of the polynomial residues is {\it process-independent}
and can be determined once and for all from the structure of the corresponding
multiple cut.  The actual value of the coefficients which appear in the residues is instead {\it process-dependent} and,  in the framework of the integrand-reduction
their determination is achieved by {\it polynomial fitting}, through the 
evaluation of the (known) integrand at values of the
loop-momenta fulfilling the cut conditions~\cite{Mastrolia:2011pr}. 

Extensions of the integrand reduction method beyond one-loop, first proposed in~\cite{Mastrolia:2011pr,Badger:2012dp}, have been 
systematized within the mathematical framework of multivariate polynomial division and basic principles of algebraic geometry~\cite{Zhang:2012ce, Mastrolia:2012an}.
Recently they become the target of several new developments~\cite{Kleiss:2012yv, Badger:2012dv,Feng:2012bm}, thus giving birth to a new direction in the study of multi-loop amplitudes. An alternative approach to two-loop calculations based on maximal unitarity has been pursued in~\cite{Gluza:2010ws, Kosower:2011ty, Larsen:2012sx, CaronHuot:2012ab}.

\section{Automation of One-Loop Calculations}

The continuous improvement of new techniques for one-loop computations led to tremendous progress in the field of NLO QCD
corrections~\cite{AlcarazMaestre:2012vp}. Calculations of increasing complexity have been performed with improved algebraic reduction
methods based on Feynman-diagrammatic algorithms, as well as with new numerical techniques
based on the idea of reconstructing one-loop amplitudes from their unitarity cuts. These theoretical
developments found an ideal counterpart in the integrand-level reduction algorithm. 

The GoSam framework~\cite{Cullen:2011ac, francesco} combines the automated algebraic generation of $d$-dimensional unintegrated amplitudes obtained via Feynman diagrams,
with the numerical integrand-level reduction.
Concerning the reduction, GoSam allows to choose at run-time (namely without regenerating the code) among integrand-level
$d$-dimensional reduction~\cite{Ossola:2006us,Ellis:2007br}, as implemented in \samurai~\cite{Mastrolia:2010nb}, or traditional tensor reduction interfaced through 
tensorial reconstruction~\cite{Binoth:2008uq, Heinrich:2010ax}. The coefficients determined by the reduction are then multiplied by the corresponding scalar (master) integrals~\cite{scalarMI}.  
Recent examples of full NLO QCD calculations performed within this framework include  $pp \rightarrow b \bar{b} b \bar{b}$~\cite{Greiner:2011mp} and $pp \rightarrow W^+\, W^- + 2$\,jet  ~\cite{Greiner:2012im} at the LHC. 

\section{Integrand-Reduction via Laurent Expansion}

An improved version of  integrand-reduction method for one-loop amplitudes was presented in \cite{Mastrolia:2012bu}, elaborating on the the techniques
proposed in~\cite{Forde:2007mi,Badger:2008cm}.
This method allows, whenever the analytic form of the numerator is
known,
to extract the unknown coefficients of the integrand decomposition by
performing a Laurent expansion.

In general, the multiple-cut conditions constrain the loop momentum.
Therefore, the on-shell solutions are parametrized by those components 
which are not completely determined in terms of the external kinematics.

The original reduction algorithm~\cite{Ossola:2007ax, Mastrolia:2008jb, Mastrolia:2010nb} requires, in these cases:
to sample the numerator on a finite subset solutions;
to subtract from the integrand all the non-vanishing contributions coming from
higher-point residues;
and finally to solve a linear system of equations in
order to find the value of the unknown coefficients parametrizing the
residue of the cut.

This algorithm can be simplified by exploiting the knowledge of the
analytic expression of the integrand.  Indeed, by performing a
\emph{Laurent expansion} with respect to one of the free parameters
which appear in the solutions of the cut,
both the integrand and the subtraction terms exhibit the same polynomial behavior of the residue.  
Moreover, the contributions coming
from the subtraction terms can be implemented as \emph{corrections at
  the coefficient level}, hence replacing the subtractions at the
integrand level of the original algorithm.  The parametric form of
this corrections can be computed once and for all, in terms of a
subset of the higher-point coefficients.  With this method the number
of coefficients entering in each subtraction term is significantly
reduced.  For instance, box and pentagons do not affect at all the
computation of lower-points coefficients.

In summary, this method identifies the coefficients of a residue with
the ones of the Laurent expansion of the numerator (with respect to one
of the free components of the loop momentum which are not fixed by the
cut conditions).  The result must be corrected by a subtraction term
which is a known function of the higher point coefficients.  If either
the analytic expression of the integrand or the tensor structure of
the numerator is known, this procedure can also be implemented in a
semi-numerical algorithm.  Indeed, the coefficients of the Laurent
expansion of a rational function can be computed, either analytically
or numerically, by performing a \empty{polynomial division} between the
numerator and the denominator.  This method has been implemented in a \C++ library, and 
preliminary tests show an improvement in 
the computational performance with respect to the standard algorithm.

\section{Higher-rank integrands}
The integrand decomposition was originally developed for renormalizable gauge theories, 
where, at one-loop, the rank $r$ of the numerator cannot be greater than the number of external legs $n$.
In \cite{Mastrolia:2012bu}, we extended the decomposition to the case where the rank becomes larger than $n$. 
This extension is required, for instance, 
for applying the integrand reduction to the production of Higgs in combination with jets, in the gluon-fusion
channel via effective-gluon vertex, generated by the large top-mass limit.
As a first step along this direction, we implemented within {\samurai}  the extension of the polynomial residues and 
the corresponding additional sampling required to fit their coefficients.

\section{Integrand-Reduction for Two-Loop Scattering Amplitudes and beyond}

The first extension of the {\it integrand reduction method} beyond one-loop
was proposed in \cite{Mastrolia:2011pr}.
A key point of the higher-loop extension is the proper parametrization
of the residues at the multi-particle poles.
We define {\it irreducible scalar products} (ISP's) the set of scalar products, among the loop momenta and 
either external momenta or polarization vectors constructed out of them, which cannot be  expressed in terms of denominators.
Residues at the multi-particle poles can be written as a multivariate polynomial in the  ISP's. 
Hence, a systematic classification of the polynomial structures of the residues is mandatory.

This task has been successfully achieved in~\cite{Mastrolia:2012an}, where we have shown that the shape of the residues is uniquely determined 
by the on-shell conditions alone, without any additional constraint. 
We have derived a simple {\it integrand recurrence relation} 
that generates the required multi-particle pole decomposition for arbitrary amplitudes, 
independently of the number of loops. 
The algorithm presented in~\cite{Mastrolia:2012an} relies on 
general properties of the loop integrands
 \beq
 \mathcal{I}_{i_1\cdots i_n}  = \frac{{\cal N}_{i_1\cdots i_n}}{D_{i_1} \cdots D_{i_n}} \; . 
 \label{Eq:Igen}
 \eeq
\begin{itemize}
\item When the number $n$ of denominators $\db{i}$ is larger than the total number of the
components of the loop momenta,
the {\it weak Nullstellensatz theorem} yields the trivial reduction of
an $n$-denominator integrand in terms integrands with $(n-1)$ denominators.
\item When  $n$ is equal or less than the total number of
components of the loop momenta, we divide the numerator  ${\cal N}_{i_1\cdots i_n}$ modulo the
Gr\"obner basis of the $n$-ple cut, namely modulo a set of polynomials
vanishing  on the same on-shell solutions as the cut denominators. The {\it remainder}
of the division is the {\it residue}  $\Delta_{i_1\cdots i_n}$ of the $n$-ple cut.  The {\it quotients} 
generate integrands with $(n-1)$ denominators which should undergo the same decomposition.

This allows us to cast the each numerator ${\cal N}_{i_1\cdots i_n}$, sitting on a set of denominators $\db{i}$, in the form
\beq
{\cal N}_{i_1\cdots i_n} =
\sum_{\kappa=1}^{n}   
{\cal N}_{i_1\cdots i_{\kappa -1}i_{\kappa+1}\cdots i_n}\, \db{i_\kappa} + \Delta_{i_1\cdots i_n} \ ,
\label{Eq:Recursive}
\eeq
which inserted in the expression for the generic the  $n$-denominator integrand, provides the aforementioned recurrence relation
\beq
\mathcal{I}_{i_1\cdots i_n}  =  
 \sum_{\kappa=1}^{n}   \mathcal{I}_{i_1\cdots i_{\kappa -1} i_{\kappa+1} i_n}
+ \frac{\Delta_{i_1\cdots i_n}}{\db{i_1} \cdots  \db{i_n}}  \, .
\label{Eq:DecGen}
\eeq
%
We remark that the procedure, together with Eqs.(\ref{Eq:Recursive}) and~(\ref{Eq:DecGen}), hold for any number of loops and in all dimensions.
\item  By iteration,  we extract  
the polynomial forms of {\it all} residues.
The algorithm will stop when all cuts are exhausted, and no denominator is left,
leaving us with the complete integrand reduction formula.
\end{itemize}

In~\cite{Mastrolia:2012an}, we have also proved a theorem on the {\it maximum-cuts}, i.e. 
the cuts defined by the maximum number of on-shell conditions which can
be simultaneously satisfied by the loop momenta.   The on-shell conditions 
of a maximum cut lead to a zero-dimensional system. 
The 
{\it Finiteness Theorem} and the 
{\it Shape Lemma} ensure that the
residue at the maximum-cut is parametrized 
by $n_s$ coefficients, where $n_s$ is the number of
solutions of the multiple cut-conditions.    
This guarantees that the
corresponding residue can {\it always be reconstructed by evaluating the numerator
at the solutions of the cut}.
The maximum cut theorem generalizes at any loop the simplicity of  
the one-loop quadruple-cut \cite{Britto:2004nc,Ossola:2006us},
where the only two solutions of the cut univocally determine the two coefficients 
needed to parametrize the residue. 

As a first application, we applied the algorithm  to a generic one-loop
integrand, reproducing  the well-known $d$-dimensional integrand decomposition formula 
\cite{Ossola:2006us,Ellis:2007br}.
Very recently~\cite{Mastrolia:2012wf}, we have applied our algorithm  to the two-loop five-point planar and 
non-planar amplitudes in ${\cal N}=4$ Super Yang-Mills (SYM) and ${\cal N}=8$ Supergravity
(SUGRA). The numerators of the integrals have at most rank two 
in the integration momenta. In particular we  perform the integrand reduction  both 
semi-numerically, by polynomial fitting, and analytically.  The latter computation 
has been performed generalizing the method of integrand reduction through Laurent expansion~\cite{Mastrolia:2012bu} discussed above.

\section{Conclusions}

The integrand reduction methods, which use the singularity structure of the integrands  to decompose the (integrated) amplitudes in terms of Master Integrals,
already proved their effectiveness in applications at the one-loop level.
Several efforts are under way in order to extend this formalism to higher orders, starting with two-loop amplitudes.

We recently proposed a new approach for the reduction of scattering
amplitudes, based on multivariate polynomial division. This technique
yields the complete integrand decomposition for arbitrary amplitudes,
regardless of the number of loops.
We have shown that the shape of the residues is uniquely determined by the on-shell conditions, and we have derived a simple integrand recurrence relation that
generates the multi-particle pole decomposition for arbitrary multi-loop amplitudes.
We have successfully applied the new reduction algorithm to one-loop and two-loop examples.

The method is well suited for a numerical implementation, but it also allows for the extraction of full analytic
results, thus providing a universal and powerful tool for the generalization of integrand-level reduction techniques
to all orders in perturbation theory.

\section*{Acknowledgments}
P.M., T.P., and H.v.D.  are
supported by the Alexander von Humboldt Foundation, in the framework of the Sofja Kovaleskaja
Award, endowed by the German Federal Ministry of Education and Research.
The work of G.O. is  supported in part by the National Science Foundation under Grant  No.~PHY-1068550
and PSC-CUNY Award No.~65188-00~43.

\end{document}